\RequirePackage{fix-cm}
\documentclass[twocolumn,epjc3]{svjour3}
\smartqed  
\RequirePackage{graphicx}
\journalname{Eur. Phys. J. C}
\begin{document}

\title{Hawking Radiation of Topological Massive Warped-AdS$_{3}$ Black Hole Families}


\author{Ganim Gecim\thanksref{e1,addr1}, Yusuf Sucu\thanksref{e2,addr1}}

\thankstext{e1}{e-mail: ganimgecim@akdeniz.edu.tr}
\thankstext{e2}{e-mail: ysucu@akdeniz.edu.tr}


\institute{Department of Physics, Faculty of Science, Akdeniz
University, 07058 Antalya, Turkey\label{addr1}}

\date{Received: date / Accepted: date}

\maketitle

\begin{abstract}
We investigate the Dirac particles tunnelling as a radiation of
Warped AdS$_{3}$ black hole family in Topological Massive Gravity.
Using the Hamilton-Jacobi method, we discuss tunnelling probability
and Hawking temperature of the spin-1/2 particles for the black hole
and its extremal cases. We observe that the Hawking temperature of
the non-extremal black hole higher than the extremal black hole when
$\omega <\frac{2\ r_{0}}{3}$, because the non-extremal black hole
become unstable in this case.
\end{abstract}
\section{Introduction}
\label{intro} A self-consistent quantum gravity theory hasn't been
constructed yet. Therefore, the quantum mechanical properties of a
classical gravitational field is studied by the quantum mechanical
behaviour of a physical system effected from it. In particular,
thanks to the extension of standard Quantum theory to curved
spacetime, some events, such as particle creation and thermal
radiation of a black hole, can be predicted. Moreover, the black
holes as the most popular concepts of the classical gravity are just
understood by the quantum mechanical concepts. From this point of
view, the solutions of the relativistic quantum mechanical wave
equations in a gravitational background became an important tool for
getting information about its nature \cite{1,2,3}. For this reason,
the relativistic quantum mechanical wave equations in a curved
spacetime background have been extensively studied \cite{4,5,6,7}.

The nature of black holes has been understood by
thermodynamical and quantum mechanical concepts since 1970 \cite%
{8,8a,8b,8c,8d,8e}. Among these concepts, thermal radiation, known
as Hawking radiation in the literature, has been investigated as a
quantum tunnelling effect of the relativistic particles from a black
hole \cite{9,9a,9b,10,10a,11,11a,11b,12,13}. Thanks to the studies,
a black hole temperature, also called hawking temperature in the
literature, is related to the black hole's surface gravity.
Therefore, the Hawking temperature becomes an important concept to
investigate the black hole physics. Since then, in the framework of
the standard Einstein general relativity, the Hawking radiation as a
tunnelling process of the particles from various black holes has
been studied extensively in the literature in both 3+1 and 2+1
dimensional spacetimes \cite{14,15,16,17,18,19,20}. On the other
hand, Kerner and Mann extended the tunnelling process to include the
Dirac particle emission from a 3+1 dimensional black hole
\cite{12,13}. Also, Ren and Li considered the Dirac particles'
tunnelling process to investigate the Hawking radiation for the
2+1-dimensional BTZ black hole by using the tunneling method
\cite{17}. The particle tunnelling process in all these studies give
useful information about the mathematical and physical properties of
the black holes. In a similar way, the Hawking radiation is used to
discuss the properties of the black hole in the context of modified
gravitation theories \cite{17a,17b,17c,17d}. As an example, Gecim
and Sucu discussed Hawking radiations for both Dirac and scalar
particles from the New-type black hole in the framework of 2+1
dimensional New Massive Gravity theory \cite{19}. However, according
to the method, both particles probe the black hole in same way.
Also, in the context of modified gravitation theories, Qi
investigated the fermion tunnelling radiation from the static
Lifshitz black hole in 2+1 dimensional New Massive Gravity theory,
and from New Class Black Holes in 3+1 Einstein-Gauss-Bonnet Gravity
\cite{20}.

The (2+1) dimensional gravitational models provide a suitable area
to investigate the quantum effects of the gravity
\cite{21,22,23,24}. Among these, Topologically Massive Gravity as an
interesting modified three-dimensional gravitation theory is formed
by adding a Chern-Simons term to the standard Einstein-Hilbert
action \cite{25}. With this term, the gravity theory has gained both
physically and mathematically interesting properties. However, in
contrast to other gravitational theories, the graviton becomes a
massive particle \cite{26,27,28,29}.

The warped AdS$_{3}$ black hole for the solution of the Topological
massive gravity is given by the following metric \cite{30}.
\begin{eqnarray}
ds^{2}=N\left( r\right) ^{2}dt^{2}-\frac{1}{N\left( r\right)
^{2}F\left( r\right) ^{2}}dr^{2}\nonumber \\-F\left( r\right)
^{2}\left[ d\phi +N^{\phi }(r)dt\right] ^{2} \label{Equation1}
\end{eqnarray}
The abbreviations used in here are as follows;
\begin{eqnarray*}
F\left( r\right) ^{2}&=&r^{2}+4\omega r+3\omega
^{2}+\frac{r_{0}^{2}}{3}\,
\\
N\left( r\right) ^{2} &=&\frac{r^{2}-r_{0}^{2}}{F\left( r\right)
^{2}}\ ,N^{\phi }(r)=-\frac{2r+3\omega }{F\left( r\right) ^{2}}.\
\end{eqnarray*}
The Warped-AdS$_{3}$ Black holes have two horizons at $r=\mp r_{0}$.
The parameters $\omega $ and $r_{0}$ are related to mass and angular
momentum of the black hole \cite{30}. The Warped-AdS$_{3}$ Black
hole becomes extremal at $r_{0}={0}$. Additionally, in the extremal
case, the black hole has a double horizon at $r={0}$. Moreover, this
result does not depend on parameter $\omega$. In an even more
special case $(\omega =r_{0}={0}),$ the metric (\ref{Equation1}) is
reduced to the horizonless metric that is characterized as ground
state or `vacuum' of the black-hole \cite{30}.

For the metric, the surface gravity is calculated by classical
(standard) method as,
\begin{eqnarray}
\kappa=\frac{1}{2}\left[ F(r)\frac{\partial }{\partial r}\left[
N^{2}(r) \right]\right]_{r=r_{0}}\ \label{Equation2}
\end{eqnarray}
and thus,
\begin{eqnarray*}
\kappa=\sqrt{3}\left( \frac{r_{0}}{2r_{0}+3\omega }\right).
\end{eqnarray*}
The Hawking temperature, $T_{H}$, defined in terms of the surface
gravity is $T_{H}=\frac{\hbar \kappa }{2\pi }$ and, for the black
hole, it is given as follows
\begin{eqnarray*}
T_{H}=\frac{\hbar\sqrt{3}}{2\pi }\left( \frac{r_{0}}{2r_{0}+3\omega
}\right).
\end{eqnarray*}

In all studies in the literature, for the non-extremal case, the
Hawking temperature is worked out by the surface gravity method that
coincides with Hawking temperature obtained by quantum mechanical
tunnelling method. In the tunnelling method, the Cauchy integral has
a first order (simple) pole in the horizon of black hole \cite
{9,9a,9b,10,10a,11,11a,11b,12,13,14,15,16,17,17a,17b,17c,17d,18,19,20}.
Therefore, the surface gravity method is agreed with tunnelling
method in the case that the Cauchy integral has a simple pole. On
the other hand, according to (\ref{Equation2}), the surface gravity
becomes zero in the extremal black hole, hence the Hawking
temperature of the extremal black hole is zero. However, in the
context of tunnelling method, as we will show in the following
section (Section 3) in this study, the relativistic quantum
mechanical particle can be tunnelled from the singularity of the
extremal black hole.

The organization of this work is as follows. In the Section 2, we
write the Dirac equation in Warped-AdS$_{3}$ Black holes background,
and calculate the tunnelling possibility of the Dirac particle by
using the semi-classical method. Also, we find Hawking temperature.
In the Section 3, we carry out the same calculation for the extremal
case, $r_{0}={0}$ (and $\omega\neq 0$), and for the ground state
$(\omega =r_{0}={0})$ of the black hole. Finally, we evaluate and
summarize the results. In this study, we use $G_{N}=k_{B}=c={1}$.

\section{Tunnelling of Dirac particles from the Warped-AdS$_{3}$ Black Hole}
\label{Dirac}

To understand the quantum mechanical properties of the black hole,
we find the probability of tunnelling and Hawking temperature by
using the solution of the relativistic quantum mechanical wave
equation for the Dirac particles. To investigate the tunnelling of
the Dirac particles from Warped-AdS$_{3}$ Black hole, we write Dirac
equation in (2+1) dimensional spacetime in the following
representation \cite{31},
\begin{eqnarray}
\left\{ i\overline{\sigma }^{\mu }(x)\left[ \partial _{\mu }-\Gamma
_{\mu }(x)\right] \right\} \Psi (x)=\frac{m_{0}}{\hbar }\Psi (x).
\label{Equation3}
\end{eqnarray}
In this representation; Dirac spinor, $\Psi (x)$, has only two
components corresponding positive and negative energy eigenstates
that has only one spin polarization. $\overline{\sigma }^{\mu }(x)$
are the spacetime dependent Dirac matrices and they are written in
terms of constant Dirac matrices, $\overline{\sigma }^{i}$, by using
triads, $e_{(i)}^{\mu }(x),$ as follows
\begin{eqnarray}
\overline{\sigma }^{\mu }(x)=e_{(i)}^{\mu }(x)\overline{\sigma
}^{i}, \label{Equation4}
\end{eqnarray}
where $\overline{\sigma }^{i}$ are Dirac matrices in a flat
spacetime and given as
\begin{eqnarray}
\overline{\sigma }^{i}=(\overline{\sigma }^{0},\overline{\sigma
}^{1}, \overline{\sigma }^{2}), \label{Equation5}
\end{eqnarray}
with
\begin{eqnarray}
\overline{\sigma }^{0}=\sigma ^{3}\ \ ,\overline{\sigma
}^{1}=i\sigma ^{1},\ \overline{\sigma }^{2}=i\sigma ^{2},
\label{Equation6}
\end{eqnarray}
where $\sigma ^{1}$, $\sigma ^{2}$ \ and $\sigma ^{3}$ \ Pauli
matrices, and $\Gamma _{\mu }(x)$ are the spin affine connection by
the following definition,
\begin{eqnarray}
\Gamma _{\mu }(x)=\frac{1}{4}g_{\lambda \alpha }(e_{\nu ,\mu
}^{i}e_{i}^{\alpha }-\Gamma _{\nu \mu }^{\alpha })s^{\lambda \nu
}(x). \label{Equation7}
\end{eqnarray}
Here, $\Gamma _{\nu \mu }^{\alpha }$ is Christoffell symbol, and
$g_{\mu\nu}(x)$ is the spacetime dependent metric tensor and given
in term of triads as follows,
\begin{eqnarray}
g_{\mu \nu }(x)=e_{\mu }^{(i)}(x)e_{\nu }^{(j)}(x)\eta _{(i)(j)},
\label{Equation8}
\end{eqnarray}
where $\mu $ and $\nu $ are curved spacetime indices running from
$0$ to $2$. $i$ and $j$ are flat spacetime indices running from $0$
to $2$ and $\eta _{(i)(j)}$ is the metric of (2+1) dimensional
Minkowski spacetime, with signature (1,-1,-1), and $s^{\lambda \nu
}(x)$ is a spin operator given by
\begin{eqnarray}
s^{\lambda \nu }(x)=\frac{1}{2}[\overline{\sigma }^{\lambda
}(x),\overline{ \sigma }^{\nu }(x)].  \label{Equation9}
\end{eqnarray}
From Eq.(\ref{Equation1}) \ and (\ref{Equation8}), the triads of $%
e_{(i)}^{\alpha }$ are written as;
\begin{eqnarray*}
e_{(0)}^{\mu }&=&\left( \frac{1}{N},0,-\frac{N^{\phi }}{N}\right), \\
e_{(1)}^{\mu }&=&\left( 0,FN,0\right), \\
e_{(0)}^{\mu }&=&\left( 0,0,\frac{1}{F}\right).
\end{eqnarray*}

The tunnelling probability for the classically forbidden trajectory
from inside to outside of the black hole horizon is given by
\begin{eqnarray}
\Gamma =e^{-\frac{2}{\hbar }ImS} \label{Equation10}
\end{eqnarray}
where $S$ is the classical action function of a particle trajectory
\cite{12,17,33,33a,33b,33c}. Therefore, in order to discuss
tunneling probability, one needs to calculate the imaginary part of
the classical action function, $S$, in regards to the tunnelling
probability \cite{9a}. To investigate the tunnelling probability of
a Dirac particle from the black hole, we use the following ansatz
for the wave function in the Eq.(\ref{Equation3});
\begin{eqnarray}
\Psi (x)=\exp \left( \frac{i}{\hbar }S\left( t,r,\phi \right)
\right)\ \left(\begin{array}{c}A\left( t,r,\phi \right) \\B\left(
t,r,\phi \right) \\ \end{array}\right)  \label{Equation11}
\end{eqnarray}
where $A\left( t,r,\phi \right) $ and $B\left( t,r,\phi \right) $
are functions of space-time \cite{17,33}. To apply the
Hamilton-Jacobi method, we insert the Eq.(\ref{Equation11}) in the
Dirac equation given by Eq.(\ref{Equation3}). Dividing by the
exponential term and neglecting the terms with $\hbar $, we derive
the following two coupled differential equations.
\begin{eqnarray}
A\left[ {m_{0}N}\left( r\right) +\frac{\partial S}{\partial
t}-N^{\phi }\left( r\right) \frac{\partial S}{\partial \phi }\right]
\nonumber \\+B\left[ iF\left(r\right) {N}\left( r\right)
^{2}\frac{\partial S}{\partial r}+\frac{{N} \left( r\right)
}{F\left( r\right) }\frac{\partial S}{\partial \phi }\right]
&=&0  \nonumber \\
A\left[ iF\left( r\right) {N}\left( r\right) ^{2}\frac{\partial S}{\partial r%
}-\frac{{N}\left( r\right) }{F\left( r\right) }\frac{\partial
S}{\partial \phi }\right]\nonumber \\ +B\left[ {m_{0}N}\left(
r\right) -\frac{\partial S}{\partial t} +N^{\phi }\left( r\right)
\frac{\partial S}{\partial \phi }\right] &=&0. \label{Equation12}
\end{eqnarray}
These two equations have nontrivial solutions for $A\left( t,r,\phi
\right)$ and $B\left( t,r,\phi \right) $ when the determinant of the
coefficient matrix is vanished. Accordingly,
\begin{eqnarray}
F\left( r\right) \left( \frac{\partial S}{\partial t}\right)
^{2}-2F\left(r\right) ^{2}N^{\phi }\left( r\right) \left(
\frac{\partial S}{\partial t} \right) \left( \frac{\partial
S}{\partial \phi }\right)\nonumber \\ +\left( F\left( r\right)
^{2}N^{\phi }\left( r\right) ^{2}-N\left( r\right) ^{2}\right)
\left( \frac{\partial S}{\partial \phi }\right) ^{2}
\nonumber \\
-N\left( r\right) ^{4}F\left( r\right) ^{4}\left( \frac{\partial
S}{\partial r}\right) ^{2}-N\left( r\right) ^{2}F\left( r\right)
^{2}m_{0}^{2}=0. \label{Equation13}
\end{eqnarray}
As $\left( \partial _{t}\right) $ and $\left( \partial _{\phi
}\right) $ are two killing vectors, we can separate $S\left(
t,r,\phi \right) $ to the variables as follows
\begin{eqnarray}
S\left( t,r,\phi \right) =-Et+j\phi +K\left( r\right) +C,
\label{Equation14}
\end{eqnarray}
where $E$ and $j$ are the energy and angular momentum of a Dirac
particle, respectively, and $C$ is a complex constant. Inserting
Eq.(\ref{Equation14}) in Eq.(\ref{Equation13}) and solving for the
radial function, $K\left(r\right) ,$ for fixed $\phi =\phi _{0}$ we
get
\begin{eqnarray}
K_{\pm }\left( r\right)=\pm \int \frac{\sqrt{F\left( r\right)
^{2}E^{2}-F\left( r\right) ^{2}N\left( r\right)
^{2}m_{0}^{2}}}{F\left( r\right) ^{2}N\left( r\right) ^{2}}dr.
\label{Equation15}
\end{eqnarray}
Because of the tunnelling event occuring at outer horizon, the
Eq.(\ref{Equation15}) can be taken as a counter integral around the
horizon to calculate the imaginary part \cite{9a}. To do this, near
the outer horizon, we can expand
$f\left(r\right)=F\left(r\right)^{2}N\left(r\right)^{2}$ as
\begin{equation}
f\left(r_{0}\right)
=\left(r-r_{0}\right)\left(\frac{df\left(r\right)}{dr}\right)
+\frac{1}{2}\left(r-r_{0}\right)^{2}\left(\frac{d^{2}f\left(
r\right) }{dr^{2}}\right) +... \label{Equation15a}
\end{equation}
Choosing the lowest-order term in Eq.(\ref{Equation15a}), the
integration in Eq.(\ref{Equation15}) around the simple pole
$r=r_{0}$ can be calculated with help of residue theorem. So we get,
\begin{equation}
K_{\pm }\left( r\right)=\pm i\pi \sqrt{3}E\left(2r_{0}+3\omega
\right) \label{Equation1AA}
\end{equation}
where $K_{+}\left( r\right) $ is outgoing and $K_{-}\left( r\right)
$ is incoming solutions of radial part. The total imaginary part of
the action is $ImS\left(t,r,\phi \right) =ImK_{\pm }\left( r\right)
=ImK_{+}\left( r\right) -ImK_{-}\left( r\right) $ \cite{19,35}.
Hence, the two kind probabilities of crossing the outer horizon,
from outside to inside or from inside to outside, are given by
\begin{eqnarray}
P_{out}=\exp \left[ -\frac{2}{\hbar }ImK_{+}\left( r\right) \right]
\nonumber \\
P_{in}=\exp \left[ -\frac{2}{\hbar }ImK_{-}\left( r\right) \right] .
\label{Equation16}
\end{eqnarray}
From the Eq.(\ref{Equation1AA}), we find that $ImK_{+}\left(
r\right)=-ImK_{-}\left( r\right) $. And, the tunnelling probability
of the Dirac particle from the outer event horizon is given by
\cite{12,33,34},
\begin{eqnarray}
\Gamma&=&\frac{P_{out}}{P_{in}}
\nonumber \\
&=&\exp \left[ -\frac{%
2\pi E\left( 2r_{0}+3\omega \right) }{\hbar \sqrt{3}r_{0}}\right].
\label{Equation17}
\end{eqnarray}
If one expands the classical action in terms of the particle energy,
the Hawking temperature is obtained at the lowest order (linear
order). So, we can write
\begin{eqnarray}
\Gamma =e^{-\frac{2}{\hbar }ImS}=e^{-\beta E}  \label{Equation18}
\end{eqnarray}
where $\beta$ is the inverse temperature of the outer horizon.
Where, the Hawking temperature is given as follows
\begin{eqnarray}
T_{H}=\frac{\hbar \sqrt{3}}{2\pi }\left( \frac{r_{0}}{2r_{0}+3\omega
}\right) \label{Equation18a}
\end{eqnarray}
This result is consistent with the result of the classical gravity.
According to this relation, the Hawking temperature increases as
$\omega $ gets smaller values of $r_{0}$. However, the angular
momentum of the black hole becomes increasingly negative in case $\omega <\frac{\sqrt{5}%
\ r_{0}}{3}$ \cite{30}. For which the black hole emit gravitational
radiation and extract angular momentum from the system
\cite{37,38,39}. It means that the system lost its energy and
momentum when $\omega <\frac{\sqrt{5}\ r_{0}}{3}$. Therefore, the
system is unstable.

\section{Tunnelling of the Particles from the Extremal Warped-AdS$_{3}$ Black Hole} \label{Extremal}

Extremal black hole solutions have an important role in the
black-hole thermodynamics. The common idea in the literature is that
the Hawking temperature of an extremal black hole vanishes because
the surface gravity is zero according to the classical method.
However, extremal black holes make radiation when they have charge
\cite{36a,36b}. On the other hand, the third law of black-hole
dynamics, the analogy between the thermodynamics and the Black hole
dynamics laws, point out that the surface gravity of a black hole
can not reach to zero \cite{36}. With this motivation, we want to
study the tunnelling event in the extremal Warped-AdS$_{3}$ black
hole by using the Hamilton-Jacobi method.

Warped-AdS$_{3}$ Black hole becomes extremal in $r_{0}={0}$
($\omega\neq 0$). In this case, the black hole has a double horizon
at $r={0}$. So the coefficients in the Eq.(\ref{Equation1}) are
reduced to following abbreviations.
\begin{eqnarray}
F\left( r\right)^{2}=r^{2}+4\omega r+3\omega ^{2},\nonumber \\
N\left( r\right)^{2}=\frac{r^{2}}{F\left( r\right) ^{2}}\
,N^{\phi}(r)=-\frac{2r+3\omega}{F\left( r\right)^{2}}.
\label{Equation21}
\end{eqnarray}
In the context of tunnelling method, inserting the
Eq.(\ref{Equation21}) and Eq.(\ref{Equation14}) in the
Eq.(\ref{Equation13}) for fixed $\phi =\phi _{0} $, the radial part
of the action function becomes as follows,
\begin{eqnarray}
K_{\pm }\left( r\right)=\pm \int \frac{\sqrt{E^{2}\left(
r^{2}+4\omega r+3\omega ^{2}\right) -r^{2}m_{0}^{2}}}{r^{2}}dr.
\label{Equation22a}
\end{eqnarray}
To calculate the counter integral in Eq.(\ref{Equation22a}), we can
use an appropriate $i\epsilon$ prescription to specify the complex
contour over which the integral has to be performed around $r=0$. In
this prescription, the singularity at $r=0$ shifted to $r=\pm
i\epsilon$ where the upper sign should be chosen for the outgoing
particles and the lower sign should be chosen for the ingoing
particles \cite{9a}. Hence, the action for the outgoing particles
with help of the contour in the upper complex plane,
\begin{eqnarray}
&S_{em}=\lim_{\varepsilon \rightarrow
0}\int\limits_{-\varepsilon }^{+\varepsilon }\frac{\sqrt{E^{2}\left(
3\omega ^{2}+4\omega \left( r-i\varepsilon \right) +\left(
r-i\varepsilon \right) ^{2}\right) -m_{0}^{2}\left( r-i\varepsilon
\right) ^{2}}}{\left( r-i\varepsilon \right)^{2}}dr& \nonumber\\
&+(real\ part)\qquad \qquad \qquad \qquad \qquad \qquad \quad \; \; \; \:&
\label{Equation22b}
\end{eqnarray}

With respect to the Residue theorem and using the
Eq.(\ref{Equation15a}), the Eq.(\ref{Equation22b}) counter integral
around the singularity located at $r={i\epsilon}$ has a second order
pole and, hence becomes,
\begin{eqnarray*}
S_{em}=(real \ part)+i\frac{2\pi E}{\sqrt{3}}.
\end{eqnarray*}
Similarly, the actions for the ingoing particles with the help of
the contour in the lower complex plane,
\begin{eqnarray}
&S_{ab}=-\lim_{\varepsilon \rightarrow
0}\int\limits_{+\varepsilon }^{-\varepsilon }\frac{\sqrt{E^{2}\left(
3\omega ^{2}+4\omega \left( r+i\varepsilon \right) +\left(
r+i\varepsilon \right) ^{2}\right) -m_{0}^{2}\left( r+i\varepsilon
\right) ^{2}}}{\left( r+i\varepsilon \right)^{2}}dr& \nonumber\\
&+(real\ part) \qquad \qquad \qquad \qquad \qquad \qquad \qquad \; \; \; \:&
\label{Equation22c}
\end{eqnarray}
and hence,
\begin{eqnarray*}
S_{ab}=(real \ part)-i\frac{2\pi E}{\sqrt{3}}.
\end{eqnarray*}
According to Eq.(\ref{Equation10}), the real part of the action does
not contribute to tunnelling probability \cite{9a}. Hence, the Dirac
particle tunnelling probability and Hawking temperature are obtained
as follows, respectively,
\begin{eqnarray*}
\Gamma =\exp \left[-\frac{8\pi E}{\hbar \sqrt{3}}\right]\
\end{eqnarray*}
and
\begin{eqnarray}
T_{H}=\frac{\hbar \sqrt{3}}{8\pi}\ \label{Equation22e},
\end{eqnarray}
where all of the contribution to the tunnelling probability and the
temperature stem from the second order pole of the complex integral
in Hamilton-Jacobi method. As we have shown above, according to
Hamilton-Jacobi method, the particles could be tunnelled from the
singularity. The Hawking temperature of this radiation becomes a
constant and, as the result of this, the extremal block hole has a
surface gravity of $\kappa =\frac{\sqrt{3}}{4}$. These results are
in agreement in regards to the analogy between the thermodynamics
laws and the Black hole dynamics laws \cite{36}.

Based on the Eq.(\ref{Equation18a}) and Eq.(\ref{Equation22e}), it
is seen that the temperature of extremal black hole is higher than
the temperature of non-extremal black hole, for all cases where
$\omega >\frac{2\ r_{0}}{3}$. However, in the $\omega <\frac{2\
r_{0}}{3}$ cases, it seems that the temperature of non-extremal
black hole is higher than the temperature of extremal black hole. On
the other hand, the super angular momentum of the non-extremal black
hole becomes increasingly negative for $\omega <\frac{2\ r_{0}}{3}$
\cite{30}. Therefore, in this case the black hole is unstable.
Hence, according to Chandrasekhar-Friedman-Schutz $(CFS)$ mechanism
\cite{37,38}, the non-extremal black hole losses its angular
momentum and energy via gravitational radiation \cite{39}. However,
the gravitational radiation gives a contribution to the Hawking
temperature. Therefore, this situation can be reasonable for the
non-extremal black hole Hawking temperature higher than the extremal
black hole Hawking temperature.

For the vacuum state $(\omega=r_{0}={0})$, the coefficients of the
metric given by Eq.(\ref{Equation1}) reduced to following
abbreviations \cite{30}:
\begin{eqnarray}
F\left( r\right)^{2}=r^{2},\ \ N\left(r\right)^{2}=1, \ \
N^{\phi}(r)=-\frac{2}{r}. \label{Equation22f}
\end{eqnarray}
If we apply the Hamilton-Jacobi method to this case, we find two
different Hawking temperatures corresponding to angular part of the
action of the tunnelling particle. Firstly, if we take arbitrary
angle, i.e. arbitrary $\phi$, and with help of the
Eq.(\ref{Equation14}), Eq.(\ref{Equation15}) and
Eq.(\ref{Equation22f}), the tunnelling probability and the Hawking
temperature of the Dirac particle can be obtained as,
\begin{eqnarray*}
\Gamma =\exp \left[-\frac{8\pi E}{\hbar \sqrt{3}}\right]\
\end{eqnarray*}
and
\begin{eqnarray*}
T_{H}=\frac{\hbar \sqrt{3}}{8\pi},
\end{eqnarray*}
which is the same result as of the $r_{0}=0$ extremal case, because
the counter integral has a second order pole in the singularity
$r=0$. Secondly, when we consider a fixed angle, i.e.
$\phi=\phi_{0}$, the counter integral which obtained by using the
Eq.(\ref{Equation14}), Eq.(\ref{Equation15}) and
Eq.(\ref{Equation22f}) gives the following equation for $W(r)$ (the
radial part):
\begin{eqnarray*}
W=\pm \int\frac{\widetilde{E}}{r}dr
\end{eqnarray*}
where $\widetilde{E}=\sqrt{E^{2}-m_{0}^{2}}$. This counter integral
has a simple pole in the singularity $r=0$. Hence, the tunnelling
probability and the Hawking temperature of the Dirac particle are
calculated as follow respectively,
\begin{eqnarray*}
\Gamma =\exp \left[-\frac{4\pi E}{\hbar}\right]\
\end{eqnarray*}
and
\begin{eqnarray*}
T_{H}=\frac{\hbar}{4\pi}.
\end{eqnarray*}
These results shown that the temperature of the vacuum state is
higher than other states.

Under the framework of findings obtained in this study, one can say
that the thermodynamical properties of the extremal case is not a
limit for thermodynamical properties the non-extremal case.

\section{Conclusions} \label{conclusions}

In this study, we have studied Hawking radiation of Dirac particles
as a quantum tunnelling effect from the Warped-AdS$_{3}$ Black
holes. By using Hamilton-Jacobi method, we have derived the
tunneling probability of the relativistic particles from the
Warped-AdS$_{3}$ Black holes. Subsequently, using the obtained these
particle tunnelling probabilities, we have calculated the Hawking
temperature for the black hole. These results are consistent with
surface gravity method based on previous works in the literature.

We have also examined the particle tunneling whether it is possible
for the extremal cases of the black hole or not. Our results show
that Dirac particles may radiate from the extremal Warped-AdS$_{3}$
black holes.

We can summarize our results as follow.
\begin{itemize}
\item ($r_{0}=0$ and $\omega\neq 0$): Hawking temperature of the
radiation from extremal Warped-AdS$_{3}$ black hole is
$T_{H}=\frac{\hbar\sqrt{3}}{8\pi}$. We infer from the quantum
mechanical result that the extremal black hole has a surface gravity
which is, classically, not predicted. From the Hawking temperature
and surface gravity relation, we get the surface gravity, as $\kappa
=\frac{\sqrt{3}}{4}$. The $\kappa$ expression can be interpreted as
quantized ground state surface gravity of the extremal
Warped-AdS$_{3}$ black hole.

\item ($r_{0}=0$ and $\omega=0$): This case represents the
vacuum state of the Warped-AdS$_{3}$ black hole. If we take an
arbitrary angle, the tunnelling probability and the Hawking
temperature of the Dirac particle can be obtained as in Case-1. But,
if we take a fixed angle, $\phi=\phi_{0}$, the tunnelling
probability and the Hawking temperature of the Dirac particle
calculated as $\Gamma=\exp \left[-\frac{4\pi E}{\hbar}\right]\ $ and
$T_{H} =\frac{\hbar}{4\pi}\ $, respectively.

\item According to $CFS$ mechanism, in the $\omega <\frac{2\
r_{0}}{3}$ case the non-extremal black hole losses its angular
momentum and energy because of gravitational radiation. Hence the
Hawking temperature of the non-extremal black hole in the $\omega
<\frac{2\ r_{0}}{3}$ case, becomes higher than the Hawking
temperature of the extremal black hole because gravitational
radiation give contribution to Hawking radiation because of the
unstability. On the other hand, temperature of extremal black hole
is higher than the temperature of non-extremal black hole, for all
cases where $\omega
>\frac{2\ r_{0}}{3}$.
\end{itemize}

All of these results show that the classical surface gravity is in
accordance with Hawking temperature calculated from the imaginary
part of the complex integral with the first order pole in the
Hamilton-Jacobi method. The classical surface gravity becomes zero
in that case the complex integral pole is of second order, but,
quantum mechanically, the particles keep tunneling from the extremal
black holes, i.e. the extremal black hole has a Hawking temperature
and thus has a surface gravity. Also, although the Hamilton-Jacobi
method predicts a surface gravity for an extremal black hole, this
method, unfortunately, terminates the spin effect of the particles
to the results. Therefore, it is obtained the same results for
scalar particle and particles with spin \cite{13,19}.

Using the Hamilton-Jacobi tunnelling method for the extremal
Reissner-Nordström (RN) black hole, one can sees that the Hawking
radiation is different zero ($T_{H}=\frac{\hbar }{4\pi M}$) in the
charged case, but the Hawking radiation vanish in the neutral case.
As shown in this study, the rotating extremal black holes have
Hawking radiation as in the charged extremal black holes
\cite{36a,36b}.

\begin{acknowledgements}
The Authors thanks Dr. Timur Sahin for useful discussion. This work
was supported by Akdeniz University, Scientific Research Projects
Unit.
\end{acknowledgements}




\end{document}